\documentclass[11pt]{article} 
\usepackage{graphicx,amssymb,epsf}
\def\ltap{\raisebox{-.4ex}{\rlap{$\sim$}} \raisebox{.4ex}{$<$}}

\begin{document}

\begin{flushright}
CERN-PH-TH/2007-264\\
OSU/HEP/07-07\\
SINP/TNP/07-33\\
\end{flushright}

\vskip 5pt

\begin{center}
{\Large \bf
Universal Doublet-Singlet Higgs Couplings and phenomenology \\ at the CERN
Large Hadron Collider} \\
\vspace*{1cm}
\renewcommand{\thefootnote}{\fnsymbol{footnote}}
{\large {\sf Gautam Bhattacharyya,${}^{1,4}$\footnote{
 {\tt gautam.bhattacharyya@saha.ac.in}}}
{\sf Gustavo C.~Branco,${}^{2,4,}$\footnote{
 {\tt gbranco@ist.utl.pt}}}  and
{\sf S. Nandi ${}^{3,4,}$\footnote{
 {\tt s.nandi@okstate.edu}}}
} \\
\vspace{10pt}
{\small ${}^{1)}${\em Saha Institute of Nuclear Physics,
1/AF Bidhan Nagar, Kolkata 700064, India}

${}^{2)}${\em
Departamento de F\'{\i}sica and Centro de
F\'{\i}sica Te\'orica de Part\'{\i}culas (CFTP), \\Instituto Superior
T\'{e}cnico, Av. Rovisco Pais, P-1049-001 Lisboa, Portugal }

${}^{3)}${\em
Department of Physics and Oklahoma Center for High Energy Physics, \\
Oklahoma State University, Stillwater, OK 74078, USA }

${}^{4)}${\em
Theory Division, CERN, CH-1211, Geneva 23, Switzerland }
}

\normalsize
\end{center}

\begin{abstract}
  We consider a minimal extension of the standard model where a real, gauge
  singlet scalar field is added to the standard spectrum.  Introducing the
  Ansatz of universality of scalar couplings, we are led to a scenario which
  has a set of very distinctive and testable predictions: (i) the mixing
  between the standard model Higgs and the new state is near maximal, (ii) the
  ratio of the two Higgs mass eigenstates is fixed ($\sim \sqrt{3}$), (iii)
  the decay modes of each of the two eigenstates are standard model like.  We
  also study how electroweak precision tests constrain this scenario.  We
  predict the lighter Higgs to lie in the range of 114 and 145 GeV, and hence
  the heavier one between 198 and 250 GeV. The predictions of the model can be
  tested at the upcoming LHC.
\end{abstract}

{\em Introduction}:~ The electroweak symmetry breaking scenario of the
standard model (SM) will undergo a thorough scrutiny at the Large
Hadron Collider (LHC).  The scalar sector of the SM consists of a
single SU(2) Higgs doublet.  However, there is no fundamental reason
why there should be only one Higgs doublet. The construction of the SM
scalar sector with just one SU(2) doublet relies essentially on the
principle of minimality.  Indeed, a single complex scalar doublet is
sufficient to implement spontaneous gauge symmetry breaking which can
account for the masses of the gauge bosons as well as the fermions.
However, in most of the extensions of the SM one has a richer Higgs
structure, with more than one doublet and/or gauge singlets. The
motivation for these extensions of the SM arises from attempts to
solving some of the drawbacks of the SM, like the hierarchy problem,
the flavor puzzle, the dark matter problem, to name a few.

At present, it is not clear what is the preferred solution to any of
the above problems. Nevertheless, quite a few such attempts advocate
the inclusion of one or more gauge singlet scalars with several
virtues.  For example, in the supersymmetric extension, the addition
of a singlet chiral superfield can provide a solution to the
$\mu$-problem in MSSM \cite{GM}. Such an addition can also aliviate
the stringent limit on the Higgs mass from LEP 2, and provide a new
invisible decay channel for the SM like Higgs boson \cite{Gunion}.
Moreover, with an additional $Z_2$ symmetry, a singlet Higgs extension
can also provide a viable candidate for the dark matter of the
universe \cite{ Murayama,Barger:2007im}.  It also facilitates first
order electroweak phase transition, as needed to produce the observed
baryon asymmetry of the universe, by lowering the Higgs mass
\cite{Profumo}.  Such electroweak singlets, together with the
doublets, also arise naturally in extra dimensional models of gauge,
Higgs and matter unification \cite{Gogoladze}.  From another
theoretical perspective, as noted in \cite{hidden1}, the Higgs mass
term in the SM is the only super-renormalizable interaction which
respects Lorentz and the SM gauge symmetry. More specifically, a
hidden sector (gauge singlet) dimension-2 scalar operator $\cal{O}$
can have an interaction like $|H|^2 {\cal{O}}^2$ which is not
suppressed by inverse powers of large scales. This leads to a mixing
between the Higgs and a gauge singlet field after $H$ acquires a
vacuum expectation value (vev).

One drawback of the general extension of the SM with a scalar singlet,
even with a $Z_2$ symmetry, is that the model is not predictive. The
masses of both the Higgs bosons, as well their mixing angle are
unknown.  The phenomenological implications for such a general extension
has been studied from the collider and cosmological perspectives
\cite{Barger:2007im,Profumo,Schabinger:2005ei}.  In the present work,
we advocate a more predictive scenario in which the only unknown
parameter is the mass of the lighter Higgs boson, and the predictions
can be easily tested at the LHC. To achieve this, for the scenario
with a real singlet scalar in addition to the SM doublet, we introduce
an Ansatz of universality of scalar couplings (as will be discussed
below), and explore its consequences within our restricted and
predictive scenario.

{\em Our model}:~ We denote the SM Higgs doublet by $H$ and the real
gauge singlet by $S$. We introduce a $Z_2$ symmetry under which $S \to
- S$. Thus $S$ is blind to both gauge and Yukawa interactions. The
scalar potential reads $(H \equiv \frac{1}{\sqrt{2}} \left(h_1+ih_2,
h_3+ih_4)^T \right)$ \ :
\begin{eqnarray}
\label{pot}
V(H,S) & = & -\mu_H^2 (H^\dagger H) - \frac{1}{2} \mu_S^2 S^2 +
\lambda_H (H^\dagger H)^2 + \frac{1}{2} \lambda_{HS} (H^\dagger H) S^2 +
\frac{1}{4} \lambda_S S^4 \ .
\end{eqnarray}
By using the minimization equations when both $H$ an $S$ acquire vevs,
and working in the unitary gauge, one obtains \ :
\begin{eqnarray}
&&H(x)  =  \frac{1}{\sqrt{2}}
\left(\begin{array}{c}
0 \\ v + h(x)
\end{array} \right) ~~, ~~
S(x) =  \eta + s(x) \ , \nonumber \\
&{\rm where} , & ~~~  v^2 = \frac{\mu_H^2 -
\frac{\lambda_{HS}}{{2\lambda_S}} \mu_S^2}
{\lambda_H - \frac{\lambda_{HS}^2}{4\lambda_S}} ~~, ~~
\eta^2 = \frac{\mu_S^2 -
\frac{\lambda_{HS}}{{2\lambda_H}} \mu_H^2}
{\lambda_S - \frac{\lambda_{HS}^2}{4\lambda_H}} \ .
\label{veta}
\end{eqnarray}
The scalar mass-squared matrix in the $(h, s)$ basis is then given by
\begin{eqnarray}
{\cal{M}}^2 = \left(
\begin{array}{cc}
2\lambda_H \, v^2 &  \lambda_{HS} \, v \, \eta \\
 \lambda_{HS} \, v \, \eta & 2\lambda_S \, \eta^2
\end{array}
\right) \ .
\label{m-sq}
\end{eqnarray}
Thus the scalar sector is described in terms of 5 parameters: the two
vevs and three dimensionless couplings, of which only the doublet vev
is known as $v \simeq 246$ GeV. The remaining 4 parameters can be
experimentally determined if both the physical scalar states are
discovered, their mixing angle is measured, and the coupling of the two
states is determined (from a possible decay of the heavier state into
the lighter one).

At this point, we introduce a simple Ansatz of ``universality of scalar
couplings (USC)'' at the weak scale, leading to
\begin{equation}
\label{univ}
\mu_H^2 \simeq \mu_S^2 \equiv \mu^2~~~
{\rm and} ~~~ \lambda_H \simeq \lambda_{HS} \simeq
\lambda_S \equiv \lambda\, .
\end{equation}

The USC Ansatz (which puts a given real component inside the doublet
$H$, e.g. $h_3$, and the real singlet $S$ at par in terms of their
strengths) is admittedly {\em ad hoc}.  Its merit resides in its
simplicity and predictive power. However, it should be noted that an
analogous hypothesis of universality of strength of Yukawa couplings
is in remarkably good agreement with the observed pattern of quark
masses and mixings \cite{USY}.  At this point one may ask whether the
USC Ansatz could result from an extra symmetry imposed on the
Lagrangian. We will comment further on this towards the end.  With
this Ansatz, the scenario becomes very predictive:
\begin{itemize}
\item The mixing angle between $h$ and $s$ is near-maximal ($\simeq
  \pi/4$).
\item The physical mass-squares are: $m_1^2 \simeq \lambda v^2$ and
  $m_2^2 \simeq 3 m_1^2$, where $v^2 \simeq 2\mu^2/3\lambda = (246
  ~{\rm GeV})^2$.
\end{itemize}
So we have only one unknown parameter, which we may take to be mass of
the lighter Higgs. It is not possible to pin down its value but we can
constrain it from direct searches and electroweak precision tests.

{\em Direct search limits}:~ The lower limit $m_h > 114.4$ GeV in the
SM arises from nonobservation of Higgs in associated production in
LEP-2 via the Bjorken process $e^+ e^- \to {Zh}$. In many extensions of
the SM (such as ours), the ZZh coupling strength for the lightest
Higgs boson is reduced by a factor $\zeta$ compared to that in the SM.
In general, the direct lower bound on the lightest Higgs boson mass is
significantly diluted if $\zeta < 0.2$ \cite{direct bound}.  In our
model, $\zeta^2 = 0.5$, and thus the SM direct mass limit of $114.4$
GeV still stands for the lighter Higgs boson in our model.  Hence the
heavier Higgs boson mass $m_2 (=\sqrt{3} m_1) > 198$ GeV.

{\em Electroweak precision tests}:~ The electroweak precision
observables significantly restrict the allowed range of masses for the
two Higgs bosons in our model. The preferred value of the SM Higgs is
$m_h = 76 ^{+33}_{-24}$ GeV. The 95\% CL upper limit of $m_h < 144$
GeV is raised to 186 GeV if the direct search limit of $m_h > 114.4$
GeV is enforced in the fit \cite{lepewwg}. The Higgs contribution to
the $T$ and $S$ parameters are logarithmic. Since the top quark mass
is now known to a very good accuracy (170.9 $\pm$ 1.8 GeV), instead of
doing a rigorous numerical fit to our model, we demand that the
two-state $\{m_1,m_2\}$ system mimics the effect of pure SM Higgs of
mass $m_h$.  If we apply this criterion on the $T$ parameter, then
using the explicit formula given in \cite{Profumo}, we obtain the
following simplified relation for maximal mixing between the two
states:
\begin{eqnarray}
\label{t}
h^{\frac{2h}{h-1}} & \simeq & {r_1}^{\frac{r_1}{r_1-1}}
{r_2}^{\frac{r_2}{r_2-1}} ,
\end{eqnarray}
where, $h = m_h^2/M_V^2$ and $r_{1,2} = m_{1,2}^2/M_V^2$. Here, $r_2=
3 r_1$ (as noted above). We take $M_V$ as an ``average'' between $W$ and
$Z$ boson masses. Using Eq.~(\ref{t}), we obtain $m_1 < 145$ GeV and
$m_2 < 250$ GeV, since $m_h < 186$ GeV .  The constraint from the $S$
parameter does not significantly alter those upper limits.

{\em Unitarity bounds}: Based on a partial wave analysis of
longitudinal gauge boson scattering, the unitarity upper limit of the
SM Higgs mass was derived as $m_h^2 ~\ltap~ 16 \pi v^2/3 \simeq (1
~{\rm TeV})^2$ \cite{unitarity}. This bound means that if the Higgs
mass exceeds the above critical value, then weak interactions will
become strong in the TeV scale and perturbation theory will break
down. In the present situation, $m_h^2$ will be replaced by
($\cos^2\phi ~ m_1^2 + \sin^2\phi ~ m_2^2$), where $\phi$ is the Higgs
mixing angle. Since $\phi \simeq \pi/4$ and $m_2^2 \simeq 3 m_1^2$, it
follows that the unitarity upper limit on the lighter Higgs state is
stronger than the SM upper limit, namely, $m_1 ~\ltap ~(1/\sqrt{2})$
TeV.

{\em Phenomenological implications}:~ Our model has several definite
phenomenological implications which can be tested at the LHC.

($i$) The model has only one unknown parameter which can be taken to
be $m_1$, the mass of the lighter Higgs boson ($h_1$). It predicts the
existence of a second Higgs boson, $h_2$, with mass $\sqrt{3}m_1$.

($ii$) The mixing angle is predicted to be maximal ($\simeq \pi/4$).
This can be tested by measuring the production cross sections of $h_1$
and $h_2$ at the LHC.

($iii$) In our model, since $m_2 < 2 m_1$, the decay $h_2 \rightarrow
h_1 h_1$ is not allowed.  This is in contrast to a large set of
general models, with a doublet and a singlet Higgs, where this decay
is allowed \cite{Profumo}.

($iv$) The branching ratios of $h_1$ and $h_2$ decays into fermions
would remain the same as in the SM, although the partial decay widths
will be equally affected due to the mixing. This feature can be tested
at the LHC and can be scrutinized even more accurately at the ILC.

($v$) Since the scalar mixing is maximal, both Higgses can be produced
with sizable cross sections at the LHC in the allowed mass ranges:
$114.4 ~\ltap~ m_1 ~\ltap~ 145$, $ 198 ~\ltap~ m_2 ~\ltap~ 250$ (all
in GeV).

{\em Discussion of the Ansatz}:~ At this point, we ask if our Ansatz of
universal scalar couplings may follow from some symmetry. In the framework of
the SU(2) $\times$ U(1) electroweak gauge model this is certainly not
possible, since the $H$ and $S$ fields transform differently under the gauge
symmetry. But constraints of the USC type could arise in a gauge extension of
the electroweak part of the SM where $H$ and $S$ would belong to the same
irreducible representation of the enlarged gauge group. We also point out that
demanding the Lagrangian being invariant under the exchange of the bilinears
$(H^\dagger H$) and $\frac{1}{2}S^2$ leads to $\mu_H^2 = \mu_S^2$ and
$\lambda_H = \lambda_S$. It will be interesting to understand the field
theoretic implications of such an exchange symmetry/relation.  This may occur,
for example, if the Higgs scalars happen to be composite objects (drawing
analogy from pions).  Clearly, it is the predictive power, testable at the
LHC, that makes our Ansatz worth considering. We mention at this stage that we
have assumed our Ansatz to be valid at the weak scale, since such relations
will in general be scale dependent unless they are protected by some symmetry.

{\em Deviation from universality}:~ A discussion of a possible mild deviation
from the Ansatz of strict universality, as expressed in Eq.~(\ref{univ}), is
now in order. For simplicity of presentation and ease of analytic
understanding, let us consider a scenario where $\mu^2 \equiv \mu_H^2, \mu_S^2
= \mu^2 + \Delta \mu^2$, and $\lambda \equiv \lambda_H = \lambda_S,
\lambda_{HS} = \lambda + \Delta \lambda$. We further assume that the
deviations are small, i.e. $\alpha \equiv \Delta \mu^2/\mu^2 \ll 1$ and $\beta
\equiv \Delta \lambda/\lambda \ll 1$. It immediately follows from
Eqs.~(\ref{veta}) and (\ref{m-sq}) that $\tan2\phi \simeq 1/(3\alpha)$ and
$m_2^2 \simeq 3 m_1^2 (1+ 4\beta/3)$. Indeed, it is quite instructive to
observe that under the above assumptions and to the leading approximations,
the deviation of the mixing angle ($\phi$) from maximality is sensitive to the
nonuniversality of mass-dimensional couplings, while the ratio between the
heavier and the lighter Higgs bosons is sensitive to the nonuniversality of
dimensionless couplings.

{\em Conclusions}:~ We have presented a simple extension of the Higgs sector
of the Standard model by adding a real electroweak singlet scalar field, and
made an Ansatz of universality of scalar couplings.  The model is highly
predictive: (i) the mixing between the two Higgs in the model is maximal, and
(ii) the mass of the heavier Higgs is $\sqrt{3}$ times that of the lighter
Higgs. These and other predictions of the model can be tested at the upcoming
LHC. The consequences of a possible mild deviation from a strict universality
have also been discussed.

\vskip 5pt
\noindent {\bf{Acknowledgements}:} All authors thank CERN Theory
Division for a very warm hospitality and support during their
sabbatical when this project was initiated. GB and SN thank the CERN
Scientific Associates Program for financial support during their stay
at CERN. We thank P.B. Pal and A. Raychaudhuri for their extremely
valuable remarks on the paper, and J. Wells for interesting
discussions. The work of GB is partially supported by the project
No.~2007/37/9/BRNS of BRNS (DAE), India.  The work of GCB was
partially supported by Funda\c c\~ ao para a Ci\~ encia e a Tecnologia
(FCT, Portugal) through projects PDCT/FP/63914/2005,
PDCT/FP/63912/2005, POCTI/FNU/44409/2002 and CFTP-FCTUNIT 777, which
are partially funded through POCTI (FEDER).  The work of SN was
supported in part by the US Department of Energy, Grant Numbers
DE-FG02-04ER41306 and DE-FG02-04ER46140.

\end{document}